\documentclass[11pt,twoside]{article}
\usepackage{amssymb}
\usepackage{latexsym}
\usepackage{graphicx}
\usepackage{amsmath}
\usepackage{hyperref,amsthm}
\usepackage{epsfig}
\usepackage{booktabs}
\usepackage{amsmath}

\numberwithin{equation}{section}
\normalsize \makeatletter
\newtheorem{thm}{Theorem}[section]
\newtheorem{lemma}[thm]{Lemma}

\newtheorem{definition}[thm]{Definition}

\newtheorem{rem}[thm]{Remark}

\newcommand{\be}{\begin{equation}}
\newcommand{\ee}{\end{equation}}
\newcommand{\bea}{\begin{eqnarray*}}
\newcommand{\eea}{\end{eqnarray*}}
\newcommand{\mR}{\mathbb{R}}

\newcommand{\mC}{\mathbb{C}}
\newcommand{\mN}{\mathbb{N}}
\DeclareMathOperator*{\argmin}{arg\,min}

\newcommand{\la}{\langle}
\newcommand{\ra}{\rangle}

\newenvironment{keyword}{\smallskip\noindent{\bf Keywords.}
                          \hskip\labelsep}{}

\begin{document}
\date{}
\title{ \textbf {Sparse Recovery with Coherent Tight Frames via Analysis Dantzig Selector and Analysis LASSO \footnote{This work is supported by NSF of China
under grant numbers 11171299 and Zhejiang Provincial NSF of China
under grant number Y6090091.}}}
\author{Junhong Lin and Song Li \thanks{Corresponding author: Song Li.
\newline E-mail adress: jhlin5@hotmail.com (J. Lin), songli@zju.edu.cn (S. Li).
\newline 2010 Mathematics Subiect Classification. Primary 94A12, 94A15, 94A08, 68P30; Secondary 41A63, 15B52, 42C15.}
\\Department of Mathematics, Zhejiang University \\Hangzhou, 310027, P. R. China }
\maketitle \baselineskip 15pt
\begin{abstract}
This article considers recovery of signals that are sparse or
approximately sparse in terms of a (possibly) highly overcomplete
and coherent tight frame from undersampled data corrupted with
additive noise. We show that the properly constrained $l_1$-analysis
optimization problem, called analysis Dantzig selector, stably
recovers a signal which is nearly sparse in terms of a tight frame
provided that the measurement matrix satisfies a restricted isometry
property adapted to the tight frame. As a special case, we consider
the Gaussian noise. Further, under a sparsity scenario, with high
probability, the recovery error from noisy data is within a log-like
factor of the minimax risk over the class of vectors which are at
most $s$ sparse in terms of the tight frame. Similar results for the
analysis LASSO are shown.

The above two algorithms provide guarantees only for noise that is
bounded or bounded with high probability (for example, Gaussian
noise). However, when the underlying measurements are corrupted by
sparse noise, these algorithms perform suboptimally. We demonstrate
robust methods for reconstructing signals that are nearly sparse in
terms of a tight frame in the presence of bounded noise combined
with sparse noise. The analysis in this paper is based on the
restricted isometry property adapted to a tight frame, which is a
natural extension to the standard restricted isometry property.

\end{abstract}

\section{Introduction}
\subsection{Standard compressed sensing }
Compressed sensing predicts that sparse signals can be reconstructed
from what was previously believed to be incomplete information. The
seminal papers \cite{CRT06a,CRT06b,D06a} have triggered a large
research activity in mathematics, engineering and computer science
with a lot of potential applications. Formally, in compressed
sensing, one considers the following model: \be \label{e7}y=Af+z,\ee
where $A$ is a known $m\times n$ measurement matrix (with $m\ll n$)
and $z\in \mR^m$ is a vector of measurement errors. The goal is to
reconstruct the unknown signal $f$ based on $y$ and $A$. The key
idea is that the sparsity helps in isolating the original signal
under suitable conditions on $A$.

The approach for solving this problem, that probably comes first to
mind, is to search for the sparsest vector in the feasible set of
possible solutions, which leads to an $l_0$-minimization problem.
However, solving the $l_0$-minimization directly is NP-hard in
general and thus is computationally infeasible \cite{MZ93,N95}. It
is then natural to consider the method of $l_1$-minimization which
can be viewed as a convex relaxation of the $l_0$-minimization.
Three most renown recovery algorithms based on convex relaxation
proposed in the literature are: the Basis Pursuit (BP) \cite{CDS01},
the Dantzig selector (DS) \cite{CT07}, and the LASSO estimator
\cite{T96} (or Basis Pursuit Denoising \cite{CDS01}):
$$\label{BP}\mbox{(BP)}:\quad\min\limits_{\tilde{f}\in{\mR^n}}\|\tilde{f}\|_1\quad\mbox{subject
to}\quad \|A\tilde{f}-y\|_2\leq \varepsilon, $$
$$\label{DS}\mbox{(DS)}:\quad\min\limits_{\tilde{f}\in{\mR^n}}\|\tilde{f}\|_1\quad\mbox{subject
to}\quad \|A^*(A\tilde{f}-y)\|_{\infty}\leq \lambda_n\sigma, $$
$$\label{LASSo}\mbox{(LASSO)}:\quad\min\limits_{\tilde{f}\in{\mR^n}}
\frac{1}{2}\|(A\tilde{f}-y)\|_2^2+\mu_n\sigma\|\tilde{f}\|_1, $$
here $\|\cdot\|_2$ denotes the standard Euclidean norm,
$\|\cdot\|_1$ is the $l_1$-norm, $\lambda_n$ (or $\mu_n$) is a
turning parameter, and $\varepsilon$ (or $\sigma$) is a measure of
the noise level. All these three optimization programs can be
implemented efficiently using convex programming or even linear
programming.

It is now well known that the BP recovers all (approximately) $s$
sparse vectors with small or zero errors provided that the
measurement matrix $A$ satisfies a restricted isometry property
(RIP) condition $\delta_{cs}\leq \delta$ for some constants
$c,\delta>0$ and that the error bound $\|z\|_2$ is small
\cite{CT05,CRT06b,C08,CWZ10,FL09,ML11}. Similar results were
obtained for the DS and the LASSO provided that $A$ satisfies a RIP
condition $\delta_{cs}\leq \delta$ for some constants $c,\delta>0$
and that  the error bound $\|A^*z\|_{\infty}$ is small
\cite{CT07,BRT09,CWZ10}. Recall that for an $m\times n$ matrix $A$
and $s\leq n$, the RIP constant $\delta_s$ \cite{CRT06a,CT06,D06b}
is defined as the smallest number $\delta$ such that for all $s$
sparse vectors $\tilde{x}\in\mR^n$,
$$(1-\delta)\|\tilde{x}\|_2^2\leq\|A\tilde{x}\|_2^2\leq(1+\delta)\|\tilde{x}\|_2^2.$$
So far, all good constructions of matrices with the RIP use
randomness. It is well known \cite{CT06,BDDW08,MPT08,RV08} that many
types of random measurement matrices such as Gaussian matrices or
Sub-Gaussian matrices have the RIP constant $\delta_s\leq\delta$
with overwhelming probability provided that $ m\geq
C\delta^{-2}s\log (n/s).$ Up to the constant, the lower bounds for
Gelfand widths of $l_1$-balls \cite{GG84,FPRU10} show that this
dependence on $n$ and $s$ is optimal. The fast multiply partial
random Fourier matrix has the RIP constant $\delta_s\leq \delta$
with very high probability provided that $m\geq C\delta^{-2}s(\log
n)^4$ \cite{CT06,RV08,KR08}.

In many common settings it is natural to assume that the noise
vector $z\sim N(0,\sigma^2 I)$, i.e., $z$ is i.i.d. Gaussian noise,
which is of particular interest in signal processing and in
statistics. The case of Gaussian noise was first considered in
\cite{HN06}, which examined the performance of $l_0$-minimization
with noisy measurements. Since the Gaussian noise is essentially
bounded (e.g. \cite{CT07,CXZ09}), all stably recovery results
mentioned above for bounded error related to the BP, the DS and the
LASSO can be extended directly to the Gaussian noise case. While the
BP and the DS (or the Lasso) provide very similar guarantees, there
are certain circumstances where the DS is preferable since the DS
yields a bound that is adaptive to the unknown level of sparsity of
the object we try to recover and thus providing a stronger guarantee
when $s$ is small \cite{CT07}. Besides, Cand\`{e}s and Tao
\cite{CT07} established an oracle inequality for the DS. Bickel et
al. \cite{BRT09} showed that the DS and the LASSO have analogous
properties, which lead to analogous error bounds.

The above mentioned recovery algorithms provide guarantees only for
noise that is bounded or bounded with high probability. However,
these algorithms perform suboptimally when the measurement noise is
also sparse \cite{LDB09}. This can occur in practice due to shot
noise, malfunctioning hardware, transmission errors, or narrowband
interference. Several recovery techniques have been developed for
sparse noise \cite{LDB09,SKPB11,LBDB11}. We refer the readers to
\cite{LDB09,SKPB11,LBDB11} and the reference therein for more
details on sparse noise.

There are many other algorithmic approaches to compressed sensing
based on pursuit algorithms in the literature, including Orthogonal
Matching Pursuit (OMP) \cite{PRK93,DMA97}, Stagewise OMP
\cite{DTDS06}, Regularized OMP \cite{NV09}, Compressive Sampling
Matching Pursuit \cite{NT08}, Iterative Hard Thresholding
\cite{BD09}, Subspace Pursuit \cite{DM09} and many other variants.
Refer to \cite{TW10} for an overview of these pursuit methods.

\subsection{$l_1$-synthesis} For signals which are sparse in the standard
coordinate basis or sparse in terms of some other orthonormal basis,
the techniques above hold. However, in practical examples, there are
numerous signals of interest which are not sparse in an orthonormal
basis. Often, sparsity is expressed not in terms of an orthogonal
basis but in terms of an overcomplete dictionary, which means that
our signal $f\in \mR^n$ is now expressed as $f= Dx$ where $D\in
\mR^{n\times d}$ ($d\geq n$) is a redundant dictionary and $x$ is
(approximately) sparse, see e.g. \cite{CDS01,BDE09,CENR11} and the
reference therein. Examples include signal modeling in array signal
processing (oversampled array steering matrix), reflected radar and
sonar signals (Gabor frames), and images with curves (Curvelet
frames), etc.

The $l_1$-synthesis (e.g. \cite{CDS01,RSV08,EMR07}) consists in
finding the sparsest possible coefficient $\hat{x}$ by solving an
$l_1$-minimization problem (BP or LASSO) with the decoding matrix
$AD$ instead of $A$, and then reconstruct the signal by a synthesis
operation, i.e., $\hat{f}=D\hat{x}$. Empirical studies show that
$l_1$-synthesis often provides good recovery \cite{CDS01,EMR07}.
Little is known about the theoretical performance of this method. In
\cite{RSV08} recovery results were obtained where essentially
require the frame $D$ to have columns that are extremely
uncorrelated such that $AD$ satisfies the RIP condition imposed by
the standard compressed sensing assumptions. However, if $D$ is a
coherent frame, $AD$ does not generally satisfy the standard RIP
\cite{RSV08,CENR11}. Also, the mutual incoherence property (MIP)
\cite{DET06} may not apply, as it is very hard for $AD$ to satisfy
the MIP as well when $D$ is highly correlated.

\subsection{$l_1$-analysis}
An alternative to $l_1$-synthesis is $l_1$-analysis, which finds the
estimator $\hat{f}$ directly by solving an $l_1$-minimization
problem. There are two most renown analysis recovery algorithms
proposed in the literature: the analysis Basis Pursuit (ABP)
\cite{CENR11} and the analysis LASSO (ALASSO)
\cite{EMR07,TSRZK05}\footnote{Note that we use the name ABP and
ALASSO as the counterparts of BP and LASSO respectively. If $D$ is
specially the concatenation of a discrete derivative and a weighted
identity, then it is the Fused LASSO introduced in \cite{TSRZK05}.}:
\be\label{ABP}(\mbox{ABP}):\quad\hat{f}=
\argmin\limits_{\tilde{f}\in{\mR^n}}\|D^*\tilde{f}\|_1\quad\mbox{subject
to}\quad \|A\tilde{f}-y\|_2\leq \varepsilon,\ee
\be\label{ALASSO}(\mbox{ALASSO}):\quad\hat{f}^{AL}=
\argmin\limits_{\tilde{f}\in{\mR^n}}\frac{1}{2}\|(A\tilde{f}-y)\|_2^2+\mu\|D^*\tilde{f}\|_1.
\ee Here $\mu$ is a tuning parameter, and $\varepsilon$ is a measure
of the noise level.

Several works exist in the literature that are related to the
analysis model (e.g. \cite{EMR07,SF09,CENR11,ACP11,LML12,NDEG11}).
It has been shown that $l_1$-analysis and $l_1$-synthesis approaches
are exactly equivalent when $D$ is orthogonal otherwise there is a
remarkable difference between the two despite their apparent
similarity \cite{EMR07}, for example truly redundant dictionaries.
Empirical evidence of the effectiveness of the analysis approach can
be found in \cite{EMR07} for signal denoising and in \cite{SF09} for
signal and image restoration. Numerical algorithms have been
proposed to solve the ALASSO, e.g. \cite{GO09,COS09,LV11}.

More recently, Cand\`{e}s et al. \cite{CENR11} showed that the ABP
recovers a signal $\hat{f}$ with an error bound
\be\label{result(1)}\|\hat{f}-f\|_2\leq
C_0\frac{\|D^*f-(D^*f)_{[s]}\|_1}{\sqrt{s}}+C_1\varepsilon,\ee
provided that $A$ satisfies a restricted isometry property adapted
to $D$ ($D$-RIP) condition with $\delta_{2s}<0.08$, where $D$ is a
tight frame for $\mR^n$. Later, the $D$-RIP condition is improved to
$\delta_{2s}<0.493$ \cite{LL11}. Note that we denote $x_{[s]}$ to be
the vector consisting of the $s$ largest coefficients of $x\in\mR^d$
in magnitude, i.e. $x_{[s]}$ is the best $s$ sparse approximation to
the vector $x$. Following \cite{CENR11}, Liu et al. \cite{LML12}
provided a theoretical study on the error when the ABP is used in
the context of compressed sensing with general frames. Aldroubi et
al. \cite{ACP11} showed that the ABP is robust to measurement noise,
and stable with respect to perturbations of the measurement matrix
$A$ and the general frames $D$. Foucart \cite{F13} studied the ABP
algorithm under the setting that the measurement matrices are
Weibull random matrices.
 Recall that the $D$-RIP of a
measurement matrix $A$, which first appeared in \cite{CENR11} and is
a natural extension to the standard RIP, is defined as follows:
\begin{definition}[$D$-RIP] Let $D$ be an $n\times d$ matrix. A measurement
matrix $A$ is said to obey the restricted isometry property adapted
to $D$ (abbreviated as $D$-RIP) of order $s$ with constant $\delta$
if
\be\label{DRIP}(1-\delta)\|Dv\|_2^2\leq\|ADv\|_2^2\leq(1+\delta)\|Dv\|_2^2\ee
holds for all $s$ sparse vectors $v\in\mR^d$. The $D$-RIP constant
$\delta_s$ is defined as the smallest number $\delta$ such that
(\ref{DRIP}) holds for all $s$ sparse vectors $v\in\mR^d$.
\end{definition} As noted in \cite{CENR11}, using a standard covering argument as in \cite{BDDW08}
(also \cite{RSV08}), one can prove that, for any $m\times n$ matrix
$A$ obeying for any fixed
$\nu\in\mR^n$,\be\label{e(concentrated)}\mathbb{P}\left(\big|\|A\nu\|_2^2-\|\nu\|_2^2\big|\geq\delta\|\nu\|_2^2\right)\leq
c{e}^{-\gamma m\delta^2},\quad \delta\in(0,1)\ee($\gamma$, $c$ are
positive numerical constants) will satisfy the $D$-RIP
$\delta_s\leq\delta$ with overwhelming probability provided that $m
\geq C\delta^{-2} s\log(d/s)$. Many types of random matrices satisfy
(\ref{e(concentrated)}). It is now well known that matrices with
Gaussian, Sub-Gaussian, or Bernoulli entries satisfy
(\ref{e(concentrated)}) (e.g. \cite{BDDW08}). It has also been shown
\cite{MPT08} that if the rows of $A$ are independent (scaled) copies
of an isotropic  $\psi_2$ vector, then $A$ also satisfies
(\ref{e(concentrated)}). Recall that an isotropic $\psi_2$ vector
$a$ is one that satisfies for all $v$, $$\mathbb{E}|\la
a,v\ra|=\|v\|_2^2\quad\mbox{and}\quad\mbox{inf}\{t:\mathbb{E}\exp(\la
a,v\ra^2/t^2)\leq 2\}\leq\alpha \|v\|_2, $$ for some constant
$\alpha$ \cite{MPT08}. Very recently, Ward and Kramer \cite{KW}
showed that randomizing the column signs of any matrix that
satisfies the standard RIP results in a matrix which satisfies the
Johnson-Lindenstrauss lemma. Therefore, nearly all random matrix
constructions which satisfy the standard RIP compressed sensing
requirements will also satisfy the D-RIP. Consequently, partial
random Fourier matrices (or partial circulant matrices) with
randomized column signs will satisfy the $D$-RIP since these
matrices are known to satisfy the RIP.

\subsection{Motivation and contributions}

In this paper, following \cite{CENR11}, we consider recovery of
signals which are (approximately) sparse in terms of a tight frame
from undersampled data. Formally, let $D$ be an $n\times d$ $(n\leq
d)$ matrix whose $d$ columns $D_1,...,D_d$ form a tight frame for
$\mR^n$, i.e.
$$f=\sum_k\langle f,D_k\rangle D_k\quad \mbox{for all}\quad
f\in\mR^n,$$ where $\langle\cdot,\cdot\rangle$ denotes the standard
Euclidean inner product. Our objective in this paper is to
reconstruct the unknown signal $f\in\mR^n$, where $D^*f$ is sparse
or approximately sparse, from a collection of $m$ linear
measurements corrupted with additive noise (\ref{e7}). Motivated by
the DS, we propose a reconstruction by the following algorithms:
\be\label{ADS}(\mbox{ADS}):\quad\hat{f}^{ADS}=
\argmin\limits_{\tilde{f}\in{\mR^n}}\|D^*\tilde{f}\|_1\quad\mbox{subject
to}\quad \|D^*A^*(A\tilde{f}-y)\|_{\infty}\leq \lambda.\ee We call
this convex program the analysis Dantzig selector (ADS). It can be
implemented efficiently using convex programming. For the rest of
this paper, $D$ is an $n\times d$ tight frame and $\delta_s$ denotes
the $D$-RIP constant with order $s$ of the measurement matrix $A$
without special mentioning.

We first show that, the ADS recovers a signal with an error bound
\be\label{error}\|\hat{f}^{ADS}-f\|_2\leq \min_{1\leq k\leq
s}\left[C_0\sqrt{k}\lambda+
C_1\frac{\|D^*f-(D^*f)_{[k]}\|_1}{\sqrt{k}}\right]\ee provided that
$A$ satisfies the $D$-RIP with $\delta_{3s}<1/2$ and that
$\|D^*A^*z\|_{\infty}\leq \lambda$, where $C_0$ and $C_1$ are small
positive constants depending only on the $D$-RIP constant
$\delta_{3s}$. As a special case, we consider the Gaussian noise
$z\sim N(0,\sigma^2 I)$. Under a sparsity scenario in the case of
Gaussian noise, comparing the error bound derived by the ABP in the
literature, e.g \cite{CENR11,LML12}, the ADS yields a bound that is
adaptive to the unknown level of sparsity (with respect to $D$) of
the object we try to recover and thus providing a stronger guarantee
when $s$ is small. Moveover, we derive a minimax over the class of
vectors which are at most $s$ sparse in terms of $D$, which tells us
that such error bound (\ref{error}) under a sparsity scenario is in
general unimprovable if one ignores the log-like factor.

To the best of our knowledge, there are fewer results on the
performance of the ALASSO in the literature related to compressed
sensing. Our second contribution of this paper is that as that for
the ADS, we derive similar results for the ALASSO .

The ADS, the ALASSO and the ABP provide guarantees only for noise
that is bounded or bounded with high probability (for example,
Gaussian noise). However, when the underlying measurements are
corrupted by sparse noise \cite{LDB09}, such algorithms fail to
recover a close approximation of the signal. Our third contribution
of this paper is that we propose robust methods for reconstructing
signals which are nearly sparse in terms of a tight frame in the
presence of bounded noise combined with sparse (with respect to a
tight frame) noise. Namely, we want to reconstruct the unknown
signal $f\in\mR^n$, where $D^*f$ is sparse or approximately sparse,
from a collection of $m$ linear measurements \be\label{e17}
y=Af+z+e,\ee where $z$ is suitably bounded, $e$ is $s'$ sparse in
terms of $\Omega$ and $\Omega\in \mR^{m\times M}$ ($M\geq m$) is a
tight frame for $\mR^m$. Let $\Phi=[A,I]$ and $u=[f^*,e^*]^*.$
Denote
$$W=\left[\begin{array}{cc}D&0\\0&\Omega\end{array}\right].$$ Then
one has $y=\Phi u+z$ and that $W\in\mR^{(n+m)\times(d+M)}$ is a
tight frame for $\mR^{n+m}$. We propose the following three
approaches: the separation ABP (SABP), the separation ADS (SADS) and
the separation ALASSO (SALASSO):
\be\label{SABP}\mbox{(SABP):}\quad\hat{u}^{SABP}=
\argmin\limits_{\tilde{u}\in{\mR^{n+m}}}\|W^*\tilde{u}\|_1\quad\mbox{subject
to}\quad \|\Phi\tilde{u}-y\|_2\leq \varepsilon,\ee
\be\label{SADS}\mbox{(SADS):}\quad\hat{u}^{SADS}=
\argmin\limits_{\tilde{u}\in{\mR^{n+m}}}\|W^*\tilde{u}\|_1\quad\mbox{subject
to}\quad \|W^*\Phi^*(\Phi\tilde{u}-y)\|_{\infty}\leq \lambda,\ee
\be\label{SALASSO}\mbox{(SALASSO):}\quad\hat{u}^{SAL}
=\argmin\limits_{\tilde{u}\in{\mR^{n+m}}}\frac{1}{2}\|(\Phi\tilde{u}-y)\|_2^2+\mu\|W^*\tilde{u}\|_1.\ee
We will provide results on the performance of these approaches in
the case when the measurement matrix $A$ is a Gaussian matrix or
Sub-Gaussian matrix. Our analysis is based on the $W$-RIP.

We shall restrict this work to the setting of real valued signals
$f\in\mR^n$. For perspective, it is known that compressed sensing
results (\cite{C08}) such as for the BP are also valid for complex
valued signals $f\in\mC^d$, e.g., \cite{F10b}. Note also that we
have restricted to the tight frame case and that a signal being
sparse in a non-tight frame is also interesting.
\subsection{Notation}
The following notation is used throughout this paper. The set of
indices of the nonzero entries of a vector $\tilde{x}$ is called the
support of $\tilde{x}$ and denoted as supp$(\tilde{x})$. Denote
$\|x\|_0=|\mbox{supp}(x)|.$ For $n\in\mN$, denote $[n]$ to mean
$\{1,2,\cdots,n\}.$ Given an index set $T\subset [n]$ and a matrix
$A\in\mR^{m\times n}$, $T^c$ is the complement of $T$ in $[n]$,
$A_T$ is the submatrix of $A$ formed from the columns of $A$ indexed
by $T$, or the $m\times n$ matrix obtained by setting the columns of
$A$ indexed by $T^c$ to zero. Write $A^*$ to mean the conjugate
transpose of a matrix $A$, $A_T^*$ to mean $(A_T)^*$,
$\lambda_{\min}(A^*A)$ and $\lambda_{\max}(A^*A)$ to mean the
smallest and largest eigenvalues of $A^*A$, $\sigma_{\min}(A)$ and
$\sigma_{\max}(A)$ to mean the smallest and largest singular values
of $A$. $\|A\|$ is the operator norm of $A$. $\|A\|_{p,q}$ denotes
the norm of $A$ from $l_p$ to $l_q$. For $j\in [n],$ $A_j$ is the
$j$th columns of $A$. $\tilde{x}_T$ is the vector equal to
$\tilde{x}$ on $T$ and zero elsewhere or a vector of $\tilde{x}$
restricted to $T$.  $C>0$ (or $c$,  $C_0$, $C_1$) denotes a
universal constant that might be different in each occurrence.

\subsection{Organization}
This paper is organized as follows. In Section 2, we present stably
recovery results for the ADS. Similar results for the ALASSO are
given in Section 3. The performance of the SABP, the SADS and the
SALASSO are presented in Section 4. Section 5 contains the proofs of
the main results.

\section{The analysis Dantzig selector}
In this section, we consider model (\ref{e7}), where $z$ is suitably
bounded. Specially, $z$ can be Gaussian noise. We will present the
recovery result of the ADS, which only requires that $A$ satisfies
the $D$-RIP.
\begin{thm}\label{thm1}Let $D$ be an arbitrary $n\times d$ tight frame and
let $A$ be an $m\times n$ measurement matrix satisfying the $D$-RIP
with $\delta_{3s}<\frac{1}{2}$.
 Assume that $\lambda$ obeys $\|D^*A^*z\|_{\infty}\leq \lambda$. Then the solution
$\hat{f}^{ADS}$ to the ADS (\ref{ADS}) obeys
$$\label{R1}\|\hat{f}^{ADS}-f\|_2\leq \min_{1\leq k\leq s}\left[C_0\sqrt{k}\lambda +
C_1\frac{\|D^*f-(D^*f)_{[k]}\|_1}{\sqrt{k}}\right],$$ where $C_0$
and $C_1$ are small constants depending only on the $D$-RIP constant
$\delta_{3s}.$\end{thm}

The Gaussian noise is essentially bounded.
\begin{lemma}\label{lemma1}Let $D$ be an arbitrary $n\times d$ tight frame and
let $A$ be an $m\times n$ matrix satisfying the $D$-RIP with
constant $\delta_1\in(0,1).$ Then for arbitrary fixed constant
$\alpha>0$, the Gaussian error $z\sim N(0,\sigma^2 I_m)$ satisfies
\bea
\mathbb{P}\left(\|D^*A^*z\|_{\infty}\leq\sigma\sqrt{2(1+\alpha)(1+\delta_1)\log
d}\right)\geq 1-\frac{1}{d^{\alpha}\sqrt{(1+\alpha)\pi\log d}}.\eea
\end{lemma}
Combining Lemma \ref{lemma1} ($\alpha=1$) with Theorem \ref{thm1}
and noting that $\delta_1\leq \delta_{3s}$, we have the following
result.
\begin{thm}\label{thm2}Let $D$ be an arbitrary $n\times d$ tight frame and
let $A$ be an $m\times n$ measurement matrix satisfying the $D$-RIP
with $\delta_{3s}<\frac{1}{2}$. Assume that $z\sim N(0,\sigma^2
I_m)$ and that $\hat{f}^{ADS}$ is the solution of the ADS
(\ref{ADS}) with $\lambda=2\sigma\sqrt{2\log d}$. Then we have
$$\label{R2}\|\hat{f}^{ADS}-f\|_2\leq \min_{1\leq k\leq s}\left[C_0\sigma\sqrt{k\log d} +
C_1\frac{\|D^*f-(D^*f)_{[k]}\|_1}{\sqrt{k}}\right]$$ with
probability at least $1-{1}/({d\sqrt{2\pi\log d}}),$ where $C_0$ and
$C_1$ are small constants depending only on $\delta_{3s}.$\end{thm}

\begin{rem}
  (a) In the exactly $s$ sparse case ($\|D^*f\|_0\leq s$), the above
theorem implies \be\label{R13}\|\hat{f}^{ADS}-f\|_2^2\leq
C_0\cdot\log d\cdot s\sigma^2.\ee Specially, when $D=I$, that is for
the standard compressed sensing, we derive similar result as in
\cite[Theorem 1.1]{CT07} (see also \cite{BRT09,CXZ09}). Now it was
shown in \cite{CT07} that the standard DS achieves a loss within a
logarithmic factor of the ideal mean squared error. The log-like
factor is the price we pay for adaptivity, that is, for not knowing
ahead of time where the nonzero coefficients actually are. In this
sense, ignoring the log-like factor, the error bound (\ref{R13}) is
in general unimprovable.

(b) The Gaussian error satisfies \be\label{GN}\mathbb{P}(\|z\|_2\leq
\sigma\sqrt{m+2\sqrt{m\log m}})\geq 1-\frac{1}{m},\ee see
\cite[Lemma 1]{CXZ09}. Combing this with (\ref{result(1)}), one
would show that the solution $\hat{f}$ to the ABP (\ref{ABP}) with
$\varepsilon =\sigma\sqrt{m+2\sqrt{m\log m}}$ satisfies
\be\label{result(2)}\|\hat{f}-f\|_2\leq
C_2\frac{\|D^*f-(D^*f)_{[s]}\|_1}{\sqrt{s}}+C_3\sigma\sqrt{m+2\sqrt{m\log
m}}\ee with high probability provided that $A$ satisfies the $D$-RIP
with $\delta_{2s}<0.493$, where $C_2$ and $C_3$ are small constants
depending on $\delta_{2s}$. Specially, if $\|D^*f\|_0\leq s$, then
\be\label{result(3)}\|\hat{f}-f\|_2\leq
C_1\sigma\sqrt{m+2\sqrt{m\log m}}.\ee Ignoring the $D$-RIP
condition, the precise constants and the probabilities with which
the stated bounds hold, we observe that in the case when $m = O(s
\log d)$, (\ref{result(3)}) and (\ref{R13}) appear to be essentially
the same. However, there is a subtle difference. Specially, if $m$
and $n$ are fixed and we consider the effect of varying $s$, we can
see that the ADS yields a bound that is adaptive to this change,
providing a stronger guarantee when $s$ is small, whereas the bound
in (\ref{result(3)}) does not improve as $s$ is reduced. What is
missing in \cite{CENR11,LML12} is achieved here is the adaptivity to
the unknown level of sparsity (with respect to $D$) of the object we
try to recover.

(c) Assume that the signal's transform coefficients in terms of $D$
decays like a power-law, i.e, the $j$th largest entry of the vector
$|D^*f|$ obeys \be\label{PL}|D^*f|_{j}\leq R\cdot j^{-1/p}\ee for
some positive numbers $R$ and $p\leq 1.$ Such a model is appropriate
for the wavelet frame coefficients of a piecewise smooth signal, for
example. Then with high probability, we have
$$\|\hat{f}^{ADS}-f\|_2^2\leq \min_{1\leq k\leq s}C_0\cdot\left(\sigma^2k\log d+R^2k^{-2/p+1}\right).$$
(In this case, one can also compare this bound with the error
estimates yielded by the ABP by applying (\ref{PL}) to
(\ref{result(2)}).) In the case of $D=I$, that is for the standard
compressed sensing, we derive similar result as in \cite[Theorem
1.3]{CT07}.

(d) We have not tried to optimize the $D$-RIP condition. We expect
that with a more complicated proof as in \cite{C08} or
\cite{FL09,CWZ10,F10,ML11}, one can still improve this condition.
\end{rem}

The error bound (\ref{R13}) is within a log-like factor of the
minimax risk over the class of vectors which are at most $s$ sparse
in terms of $D$:
\begin{thm}\label{thm5}
  Let $D$ be an arbitrary $n\times d$ tight frame.
  Assume that the measurement matrix $A$ satisfies the $D$-RIP of order $s$ and that $z\sim N(0,\sigma^2
  I_m).$ Suppose that there exists a subset $T_0\in [d]$ such that $|T_0|=s$
  and $\Sigma_{T_0}\subset\{D^*\tilde{f}:\tilde f\in\mR^n\},$ where $\Sigma_{T_0}=\{x\in\mR^d:\mbox{supp}(x)\subset
  T_0\}.$ Then $$\inf_{\hat{f}}\sup_{\|D^*f\|_0\leq s}\mathbb{E}\|\hat{f}-f\|_2^2\geq\frac{1}{1+\delta_s}s\cdot
  \sigma^2,$$ where the infimum is over all measurable functions
  $\hat{f}(y)$ of $y.$
\end{thm}
\begin{rem}
  When $D$ is an identity matrix or an orthonormal basis, the
  condition $\Sigma_{T_0}\subset\{D^*\tilde{f}:\tilde f\in\mR^n\}$
  is satisfied.
\end{rem}

The exacting reading may argue that while this lower bound is in
expectation, the upper bound holds with high probability. Thus, we
provide the following complementary theorem.
\begin{thm}\label{thm6}
  Under the assumptions of Theorem \ref{thm5}, any estimator $\hat{f}(y)$ obeys
  $$\sup_{\|D^*f\|_0\leq s}\mathbb{P}\left(\|\hat{f}-f\|_2^2\geq\frac{1}{2(1+\delta_s)}s\cdot \sigma^2\right)
  \geq1-e^{-\frac{s}{16}}.$$
\end{thm}

\section{The analysis LASSO}
In this section, we will present the performance of the ALASSO from
the noisy measurements  (\ref{e7}), where $z$ is suitably bounded.
Specially, $z$ can be Gaussian noise. Note that our results are
similar as that for the ADS.

\begin{thm}\label{thm3}
Let $D$ be an arbitrary $n\times d$ tight frame and let $A$ be an
$m\times n$ measurement matrix satisfying the $D$-RIP with
$\delta_{3s}<\frac{1}{4}$.
 Assume that $\mu$ obeys $\|D^*A^*z\|_{\infty}\leq \mu/2$. Then the solution
$\hat{f}^{AL}$ to the ALASSO (\ref{ALASSO}) obeys
$$\label{R3}\|\hat{f}^{AL}-f\|_2\leq \min_{1\leq k\leq s}\left[C_0\sqrt{k}\mu +
C_1\frac{\|D^*f-(D^*f)_{[k]}\|_1}{\sqrt{k}}\right],$$ where $C_1$ is
small constant depending only on $\delta_{3s}$ and $C_0$ is
depending on $\delta_{3s}$ and $\|D^*D\|_{1,1}.$ \end{thm}

Combining Lemma \ref{lemma1} with Theorem \ref{thm3}, we have the
following result.
\begin{thm}\label{thm4}
Let $D$ be an arbitrary $n\times d$ tight frame and let $A$ be an
$m\times n$ measurement matrix satisfying the $D$-RIP with
$\delta_{3s}<\frac{1}{4}$. Assume that $z\sim N(0,\sigma^2 I_m)$ and
that $\hat{f}^{AL}$ is the solution of the ALASSO  with
$\mu=4\sigma\sqrt{2\log d}$. Then we have
$$\label{R4}\|\hat{f}^{AL}-f\|_2\leq \min_{1\leq k\leq s}\left[C_0\sigma\sqrt{k\log d}+
C_1\frac{\|D^*f-(D^*f)_{[k]}\|_1}{\sqrt{k}}\right]$$ with
probability exceeding $1-{1}/({d\sqrt{2\pi\log d}}),$ where $C_1$ is
small constant depending only on $\delta_{3s}$ and $C_0$ is
depending on $\delta_{3s}$ and $\|D^*D\|_{1,1}.$ \end{thm}

\begin{rem}
 (a) From the proof, one can see that
$C_0=2\sqrt{2}(1+2\|D^*D\|_{1,1})/(1-4\delta_{3s}).$
  When $D$ is an identity matrix or an orthonormal basis,
  $\|D^*D\|_{1,1}=1.$ For general tight frame $D$, we hope that with some more delicate proof, the
  depending on $\|D^*D\|_{1,1}$ can be deleted.

  (b) In the exactly $s$ sparse case ($\|D^*f\|_0\leq s$), the above
theorem implies $$\label{R14}\|\hat{f}^{AL}-f\|_2^2\leq C_0\cdot\log
d\cdot s\sigma^2.$$ Specially, when $D=I$, that is for the standard
compressed sensing, we derive similar result as in \cite[Theorem
7.2]{BRT09}.
\end{rem}

\section{Sparse noise} In this section, we consider
model (\ref{e17}), where $z$ is suitably bounded and $e$ is sparse
in terms of a tight frame $\Omega$.
\begin{thm}\label{thm7}Let $D$ be an arbitrary $n\times d$ tight frame
and let $A$ be an $m\times n$ matrix with elements $a_{ij}$ drawn
  i.i.d according to $N(0,1/m)$. Let $\|\Omega^*e\|_0\leq s'$, where $\Omega\in\mR^{m\times M}$ is a tight frame for $\mR^m$.
  Suppose $m\geq C\delta^{-2}(s+s')\log
((d+M)/(s+s'))$ for some fixed $\delta\in(0,1/4)$ and constant $C$.

(a) Let $\lambda$ obeys $\|W^*\Phi^*z\|_{\infty}\leq \lambda$. Then
with high probability, the solution $\hat{u}^{SADS}$ to (\ref{SADS})
obeys
$$\label{R10}\|\hat{f}^{SADS}-f\|_2\leq\|\hat{u}^{SADS}-u\|_2\leq
\min_{1\leq k\leq s}\left[C_0\sqrt{k+s'}\lambda +
C_1\frac{\|D^*f-(D^*f)_{[k]}\|_1}{\sqrt{k+s'}}\right].$$

(b) Let $\mu$ obeys $\|W^*\Phi^*z\|_{\infty}\leq \mu/2$. Then with
high probability, the solution $\hat{u}^{SAL}$ to (\ref{SALASSO})
obeys
$$\label{R11}\|\hat{f}^{SAL}-f\|_2\leq\|\hat{u}^{SAL}-u\|_2\leq
\min_{1\leq k\leq s}\left[C_2(1+2\|D^*D\|_{1,1})\sqrt{k+s'}\mu +
C_3\frac{\|D^*f-(D^*f)_{[k]}\|_1}{\sqrt{k+s'}}\right].$$

(c) Assume that $\|z\|_2\leq \varepsilon.$ Then with high
probability, the solution $\hat{u}^{SABP}$ to (\ref{SABP}) obeys
$$\|\hat{f}^{SABP}-f\|_2\leq \|\hat{u}^{SABP}-u\|_2\leq
C_4\varepsilon + C_5\frac{\|D^*f-(D^*f)_{[s]}\|_1}{\sqrt{s+s'}}.$$
In the above, $C_0,\cdots,C_5$ are small constants depending only on
$\delta.$
\end{thm}

\begin{rem}
  (a) From the proof of this theorem, one can see that such results can be extended to
  the more general class of Sub-Gaussian
  matrices and the case that $e$ is nearly sparse in terms of $\Omega$.

  (b) In the case of $z=0$ and $\|D^*f\|_0\leq s$, the
above theorem implies exact recovery (both $f$ and $e$) via
$$\label{S}\hat{u}=
\argmin\limits_{\tilde{u}\in{\mR^{n+m}}}\|W^*\tilde{u}\|_1\quad\mbox{subject
to}\quad \Phi\tilde{u}=y.$$ Specially, when $\Omega=I$, we derive
similar result as in \cite{LDB09}.

  (c) By applying Lemma
  \ref{lemma1} (Since from the proof of this theorem, one can see that $A$ satisfies the $W$-RIP) and (\ref{GN}) to the above theorem, one can get error
  estimates for the SABP, the SADS and the SALASSO in the case of $z\sim N(0,\sigma^2 I).$
\end{rem}

\section{Proofs}
We first recall some useful properties of a tight frame. Let $D$ be
an arbitrary $n\times d$ tight frame for $\mR^n$, then
$$\|f\|_2^2=\|D^*f\|_2^2\quad \mbox{for all}\quad
f\in\mR^n,\quad\mbox{and}\quad \|Dv\|_2\leq \|v\|_2\quad\mbox{for
all}\quad v\in\mR^d.$$ Refer the readers to \cite[Chapter 3]{D92}
for details.
\subsection{Proof of Lemma \ref{lemma1}}
\begin{proof}[Proof of Lemma \ref{lemma1}]Note that from the definition of $D$-RIP, we have
  \be\label{e1}\sqrt{1-\delta_1}\|D_j\|_2\leq\|AD_j\|_2\leq\sqrt{1+\delta_1}\|D_j\|_2\leq \sqrt{1+\delta_1},\quad \forall j\in[d].\ee
  Without loss of generality, we assume that $\|D_j\|_2\neq 0$ for each $j\in[d]$. Then by (\ref{e1}), we have $\|AD_j\|_2\neq 0.$
  Let $\omega_j=\frac{\la AD_j,z\ra}{\sigma\|AD_j\|_2}.$ Then $\omega_j$ has Gaussian
  distribution $N(0,1).$ By using the union bound and then the inequality
  (\ref{e1}), we get
\bea &&\mathbb{P}\left(\|D^*A^*z\|_{\infty}>\sigma\sqrt{2(1+\alpha)(1+\delta_1)\log d}\right)\\
  &\leq&\sum_{j=1}^{d}\mathbb{P}\left(|\omega_j|\|AD_j\|_2>\sqrt{2(1+\alpha)(1+\delta_1)\log
  d}\right)\\
  &\leq&\sum_{j=1}^{d}\mathbb{P}\left(|\omega_j|>\sqrt{2(1+\alpha)\log d}\right)\\
  &=&d\cdot\mathbb{P}\left(|\omega_1|>\sqrt{2(1+\alpha)\log d}\right)\leq\frac{1}{d^{\alpha}\sqrt{(1+\alpha)\pi\log d}},\eea
  where the last step follows from the Gaussian tail probability bound that for a standard Gaussian variable $V$ and any constant $t$,
  $\mathbb{P}\left(|V|>t\right)\leq2t^{-1}\frac{1}{\sqrt{2\pi}}e^{-\frac{1}{2}t^2}$.
  It thus follows that \bea&&\mathbb{P}\left(\|D^*A^*z\|_{\infty}\leq\sigma\sqrt{2(1+\alpha)(1+\delta_1)\log
  d}\right)\\
  &=&1-\mathbb{P}\left(\|D^*A^*z\|_{\infty}>\sigma\sqrt{2(1+\alpha)(1+\delta_1)\log d}\right)\\&\geq&
  1-\frac{1}{d^{\alpha}\sqrt{(1+\alpha)\pi\log d}}.\eea
\end{proof}

\subsection{Proof of Theorem \ref{thm1}}
\begin{proof}[Proof of Theorem \ref{thm1}] The proof makes use of the ideas
from \cite{CENR11,CT07,C08,CP11}.
  Let $f$ and $\hat{f}^{ADS}$ be as in the theorem, and let $T_0=T$ denote the set of the $s$ largest coefficients of
  $D^*f$ in magnitude.
  Set $h=\hat{f}^{ADS}-f$ and observe that by the triangle
  inequality \be\label{DAAH}\|D^*A^*Ah\|_{\infty}\leq\|D^*A^*(Af-y)\|_{\infty}+\|D^*A^*(A\hat{f}^{ADS}-y)\|_{\infty}\leq2\lambda.\ee
  Since $\hat{f}^{ADS}$ is a minimizer, one gets that
  $$\|D^*f\|_1\geq\|D^*\hat{f}^{ADS}\|_1.$$ That is
  $$\|D^*_Tf\|_1+\|D^*_{T^c}f\|_1\geq\|D^*_T\hat{f}^{ADS}\|_1+\|D^*_{T^c}\hat{f}^{ADS}\|_1.$$
  Thus $$\|D^*_Tf\|_1+\|D^*_{T^c}f\|_1\geq\|D^*_Tf\|_1-\|D^*_Th\|_1+\|D^*_{T^c}h\|_1-\|D^*_{T^c}f\|_1.$$ This implies \be\label{CC}
  \|D^*_{T^c}h\|_1\leq2\|D^*_{T^c}f\|_1+\|D^*_Th\|_1.\ee

  Next, we decompose the coordinates $T_0^c$ into sets of size $s$ in order of decreasing magnitude of
  $D^*_{T^c}h$. Denote these sets $T_1, T_2,...,$ and for simplicity of notation set $T_{01} = T_0\cup
  T_1$. Note that for each $j\geq2,$
  $$\|D_{T_j}^*h\|_2\leq s^{1/2}\|D_{T_j}^*h\|_{\infty}\leq
  s^{-1/2}\|D_{T_{j-1}}^*h\|_{1}$$ and thus
  \be\label{e2} \sum\limits_{j\geq2}\|D_{T_j}^*h\|_2\leq \sum\limits_{j\geq1}s^{-1/2}\|D_{T_j}^*h\|_1=s^{-1/2}\|D_{T^c}^*h\|_1.\ee

  Set $u_{01}=D^*_{T_{01}}h/\|DD^*_{T_{01}}h\|_2$ and $u_j=D^*_{T_j}h/\|DD^*_{T_j}h\|_2$ for each
  $j\geq2$. Then $\|Du_{01}\|_2=1$ and $\|Du_j\|_2=1$ for each $j\geq 2$. We then obtain that
  \bea &&\frac{\la ADD^*_{T_{01}}h,ADD^*_{T_j}h\ra}{\|DD^*_{T_{01}}h\|_2\|DD^*_{T_j}h\|_2}=\langle ADu_j,ADu_{01}\rangle
  =\frac{1}{4}\left\{\|ADu_j+ADu_{01}\|_2^2-\|ADu_{j}-ADu_{01}\|_2^2\right\}\\ &\geq&
\frac{1}{4}\left\{(1-\delta_{3s})\|Du_j+Du_{01}\|_2^2-(1+\delta_{3s})\|Du_j-Du_{01}\|_2^2\right\}\\
&=& \langle
Du_j,Du_{01}\rangle-\frac{\delta_{3s}}{2}\left\{\|Du_j\|_2^2+\|Du_{01}\|_2^2\right\}=
\langle Du_j,Du_{01}\rangle-\delta_{3s}.\eea It thus follows that
  \bea &&\la Ah,ADD^*_{T_{01}}h\ra=\la ADD^*_{T_{01}}h,ADD^*_{T_{01}}h\ra+\sum_{j\geq2}\la ADD^*_{T_j}h,ADD^*_{T_{01}}h\ra\\
&\geq&(1-\delta_{3s})\|DD^*_{T_{01}}h\|_2^2-\delta_{3s}\|DD^*_{T_{01}}h\|_2\sum_{j\geq2}\|DD^*_{T_j}h\|_2
+\sum_{j\geq2}\la DD^*_{T_j}h,DD^*_{T_{01}}h\ra.\eea By applying the
equality
$$\sum_{j\geq2}\la DD^*_{T_j}h,DD^*_{T_{01}}h\ra=\la h- DD^*_{T_{01}}h,DD^*_{T_{01}}h\ra=\|D^*_{T_{01}}h\|_2^2-\|DD^*_{T_{01}}h\|_2^2,$$
we get \bea\la
Ah,ADD^*_{T_{01}}h\ra&\geq&\|D^*_{T_{01}}h\|_2^2-\delta_{3s}\|DD^*_{T_{01}}h\|_2^2-\delta_{3s}\|DD^*_{T_{01}}h\|_2\sum_{j\geq2}\|DD^*_{T_j}h\|_2\\
&\geq&(1-\delta_{3s})\|D^*_{T_{01}}h\|_2^2-\delta_{3s}\|D^*_{T_{01}}h\|_2\sum_{j\geq2}\|D^*_{T_j}h\|_2.\eea
Substituting the inequality (\ref{e2}) into the above inequality, we
derive \bea\la Ah,ADD^*_{T_{01}}h\ra
\geq(1-\delta_{3s})\|D^*_{T_{01}}h\|_2^2-s^{-1/2}\delta_{3s}\|D^*_{T_{01}}h\|_2\|D_{T^c}^*h\|_1.\eea
Besides, by using the holder inequality and (\ref{DAAH}), we have
\bea\la Ah,ADD^*_{T_{01}}h\ra=\la D^*A^*Ah,D^*_{T_{01}}h\ra\leq
\|D^*A^*Ah\|_{\infty}\|D^*_{T_{01}}h\|_1\leq 2\lambda
\sqrt{2s}\|D^*_{T_{01}}h\|_2.\eea Now combining the above two
inequalities and by an easy computation, we can derive
\be\label{e3}\|D^*_{T_{01}}h\|_2\leq
\frac{2\lambda\sqrt{2s}+s^{-1/2}\delta_{3s}\|D_{T^c}^*h\|_1}{1-\delta_{3s}}.\ee
It thus follows that $$\|D^*_{T}h\|_1\leq
\sqrt{s}\|D^*_{T}h\|_2\leq\sqrt{s}\|D^*_{T_{01}}h\|_2\leq
\frac{2\sqrt{2}\lambda
s+\delta_{3s}\|D_{T^c}^*h\|_1}{1-\delta_{3s}}.$$ Substituting the
above inequality to (\ref{CC}) and by an easy calculation, we can
obtain
\be\label{e4}\|D^*_{T^c}h\|_1\leq\frac{2(1-\delta_{3s})\|D^*_{T^c}f\|_1+2\sqrt{2}\lambda
s}{1-2\delta_{3s}}.\ee

Now we are ready to give the error estimates. Note that
\bea\|h\|_2=\|D^*h\|_2\leq\|D_{T_{01}}^*h\|_2+\sum_{j\geq2}\|D_{T_j}^*h\|_2.\eea
Introducing (\ref{e2}) and (\ref{e3}) to the above, we get
\bea\|h\|_2\leq
\frac{2\lambda\sqrt{2s}+s^{-1/2}\|D_{T^c}^*h\|_1}{1-\delta_{3s}}.\eea
By applying (\ref{e4}), we derive \bea\|h\|_2 \leq
\frac{4\sqrt{2s}\lambda
}{1-2\delta_{3s}}+\frac{2\|D^*_{T^c}f\|_1}{(1-2\delta_{3s})\sqrt{s}}.\eea

Repeating the above argument for each $1\leq k<s$, one can prove
that \bea\|h\|_2 \leq \frac{4\sqrt{2k}\lambda
}{1-2\delta_{3k}}+\frac{2\|D^*_{T^c}f\|_1}{(1-2\delta_{3k})\sqrt{k}}.\eea
Now the proof can be finished by noting that $\delta_{3k}\leq
\delta_{3s}$ for $1\leq k\leq s.$
\end{proof}

\subsection{Proof of Theorem \ref{thm5}}
We first introduce the following well-known lemma, see for example
\cite[Lemma 3.11]{CP11}. It gives the minimax risk for estimating
the vector $x\in\mR^s$ from the data $y\in\mR^m$ and the linear
model \be\label{e9}y=\Phi x+z,\ee where $\Phi\in\mR^{m\times s}$ and
$z\sim N(0,\sigma^2 I_m).$

\begin{lemma}\label{lemma4}
Let $\Phi,x,y,z$ follow the linear model (\ref{e9}) and that
$\lambda_{i}(\Phi^*\Phi)$ be the eigenvalues of the matrix
$\Phi^*\Phi.$
  Then $$\inf_{\hat{x}}\sup_{x\in\mR^s}\mathbb{E}\|\hat{x}-x\|_2^2=\sigma^2\mbox{trace}((\Phi^*\Phi)^{-1})
  =\sum_{i}\frac{\sigma^2}{\lambda_{i}(\Phi^*\Phi)},$$ where the infimum
  is over all measurable functions $\hat{x}(y)$ of $y.$ In
  particular, if one of the eigenvalues vanishes, then the minimax
  risk is unbounded.
\end{lemma}
\begin{proof}[Proof of Theorem \ref{thm5}]
  Note that we have
  \begin{eqnarray}\inf_{\hat{f}}\sup_{\|D^*f\|_0\leq s}\mathbb{E}\|\hat{f}-f\|_2^2
  &\geq& \inf_{\hat{f}}\sup_{D^*f\in\Sigma_{T_0}}\mathbb{E}\|\hat{f}-f\|_2^2\nonumber\\
  &=&\inf_{\hat{f}}\sup_{D^*f\in\Sigma_{T_0}}\mathbb{E}\|D^*\hat{f}-D^*f\|_2^2\nonumber
  \\&\geq&\inf_{\hat{f}}\sup_{D^*f\in\Sigma_{T_0}}\mathbb{E}\|D_{T_0}^*\hat{f}-D_{T_0}^*f\|_2^2\label{eq:1}.\end{eqnarray}
  For each $f$ such that $D^*f\in\Sigma_{T_0},$ we rewrite the original model
  $y=Af+z$ as $y=AD_{T_0}v+z$, where $v\in\mR^{s}$ and $z\sim N(0,\sigma^2 I_m).$
  Since we have
  $\Sigma_{T_0}\subset\{D^*\tilde{f}:\tilde{f}\in\mR^n\}$ and that $D_{T_0}^*\hat{f}(y)$ is measurable of
  $y$, we get
  \be\label{eq:2}
  \inf_{\hat{f}}\sup_{D^*f\in\Sigma_{T_0}}\mathbb{E}\|D_{T_0}^*\hat{f}-D_{T_0}^*f\|_2^2
  \geq\inf_{\hat{v}}\sup_{v\in\mR^{s}}\mathbb{E}\|\hat{v}-v\|_2^2,\ee
  where $v\in\mR^s,AD_{T_0},y,z$ follow the linear model
  $y=AD_{T_0}v+z,$ $z\sim N(0,\sigma^2 I_m),$ and the infimum of the
  last term
  is over all measurable functions $\hat{v}(y)$ of $y.$
  Note that from the definition of $D$-RIP, for all $v\in\mR^s$, we
have
$$\|AD_{T_0}v\|_2^2\leq
(1+\delta_s)\|D_{T_0}v\|_2^2\leq(1+\delta_s)\|v\|_2^2.$$
  It thus follows that \be\label{e8}\lambda_{\max}(D_{T_0}^*A^*AD_{T_0})\leq1+\delta_s.\ee
   By using Lemma
  \ref{lemma4} and (\ref{e8}), we have
  \be\label{e18}\inf_{\hat{v}}\sup_{v\in\mR^{s}}\mathbb{E}\|\hat{v}-v\|_2^2=\sum_{i}\frac{\sigma^2}{\lambda_{i}(D_{T_0}^*A^*AD_{T_0})}\geq\frac{1}{1+\delta_s}s\cdot \sigma^2.\ee
  Introducing (\ref{eq:2}) and (\ref{e18}) to (\ref{eq:1}), we
  derive
  \bea\inf_{\hat{f}}\sup_{\|D^*f\|_0\leq s}\mathbb{E}\|\hat{f}-f\|_2^2
  \geq\frac{1}{1+\delta_s}s\cdot \sigma^2.\eea
\end{proof}

\subsection{Proof of Theorem \ref{thm6}}
We begin by introducing the following lemma, see \cite[Lemma
3.14]{CP11}.
\begin{lemma}\label{lemma5}
  Suppose that $x,y,\Phi,z$ follow the linear model (\ref{e9}) with $z\sim N(0,\sigma^2
  I).$ Then
  $$\inf_{\hat{x}}\sup_{x\in\mR^s}\mathbb{P}\left(\|\hat{x}-x\|_2^2\geq\frac{1}{2\|\Phi\|^2}s\cdot\sigma^2\right)\geq
  1-e^{-\frac{s}{16}}.$$
\end{lemma}
\begin{proof}[Proof of Theorem \ref{thm6}]
  From the definition of tight frame, we have
  \bea&&\sup_{\|D^*f\|_0\leq s}\mathbb{P}\left(\|\hat{f}-f\|_2^2\geq\frac{1}{2(1+\delta_s)}s\cdot \sigma^2\right)
  \\&=&\sup_{\|D^*f\|_0\leq s}\mathbb{P}\left(\|D^*\hat{f}-D^*f\|_2^2\geq\frac{1}{2(1+\delta_s)}s\cdot
  \sigma^2\right)\\&\geq&\sup_{D^*f\in\Sigma_{T_0}}\mathbb{P}\left(\|D^*\hat{f}-D^*f\|_2^2\geq\frac{1}{2(1+\delta_s)}s\cdot
  \sigma^2\right)\\&\geq&\sup_{D^*f\in\Sigma_{T_0}}\mathbb{P}\left(\|D_{T_0}^*\hat{f}-D_{T_0}^*f\|_2^2\geq\frac{1}{2(1+\delta_s)}s\cdot
  \sigma^2\right)\\&\geq& \sup_{D^*f\in\Sigma_{T_0}}\mathbb{P}\left(\|D_{T_0}^*\hat{f}-D_{T_0}^*f\|_2^2\geq\frac{1}{2\|AD_{T_0}\|^2}s\cdot\sigma^2\right),\eea
  where we have used (\ref{e8}) for the last
  step. Note that $D_{T_0}^*\hat{f}(y)$ is measurable of $y$ since $\hat{f}(y)$ is
  measurable. Then, with the assumption
  $\Sigma_{T_0}\subset\{D^*\tilde{f}:\tilde{f}\in\mR^n\}$, we get
   \bea &&\sup_{\|D^*f\|_0\leq s}\mathbb{P}\left(\|\hat{f}-f\|_2^2\geq\frac{1}{2(1+\delta_s)}s\cdot
  \sigma^2\right)\\&\geq&\sup_{D^*f\in\Sigma_{T_0}}\mathbb{P}\left(\|D_{T_0}^*\hat{f}-D_{T_0}^*f\|_2^2\geq\frac{1}{2\|AD_{T_0}\|^2}s\cdot
  \sigma^2\right)\\
  &\geq&\inf_{\hat{v}}\sup_{v\in\mR^s}\mathbb{P}\left(\|\hat{v}-v\|_2^2\geq\frac{1}{2\|AD_{T_0}\|^2}s\cdot \sigma^2\right)\\
  &\geq&1-e^{-\frac{s}{16}},\eea where the last step follows
  from Lemma \ref{lemma5}.\end{proof}

\subsection{Proof of Theorem \ref{thm3}}
\begin{proof}[Proof of Theorem \ref{thm3}]The proof is similar to that of Theorem \ref{thm1}.
Set $h=\hat{f}^{AL}-f.$ We will prove the following two
inequalities:
\begin{itemize}
    \item {$\|D^*A^*(A\hat{f}^{AL}-y)\|_{\infty}\leq\mu \|D^*D\|_{1,1}$.}
    \item {$\|D^*_{T^c}h\|_1\leq3\|D^*_Th\|_1+4\|D^*_{T^c}f\|_1.$}
\end{itemize}
With these two inequalities and the assumptions of this theorem, a
similar approach as that for Theorem \ref{thm1} would lead to our
results.

For convenience, we denote $\mathcal{L}$ as the function
$$\mathcal{L}(\tilde{f})=\frac{1}{2}\|(A{\tilde{f}}-y)\|_2^2+\mu\|D^*{\tilde{f}}\|_1,$$ in
which $\mu=4\sigma\sqrt{2\log d}.$ The subdifferential $\partial
\mathcal{F}$ of a real valued convex lower semicontinuous function
$\mathcal{F}: \mR^n\rightarrow\mR$ is the multifunction defined by
$$\partial \mathcal{F}(f_0)=\left\{g\in\mR^n|\forall \tilde{f}\in\mR^n,\quad
\mathcal{F}(\tilde{f})\geq \mathcal{F}(f_0)+\la
g,\tilde{f}-f_0\ra\right\}.$$ Note that $f_0$ is a minimum of
$\mathcal{F}$ if and only if $0\in
\partial \mathcal{F}(f_0).$ The subdifferential
of $\mathcal{L}(\hat{f}^{AL})$ is $$\partial
\mathcal{L}(\hat{f}^{AL})=\left\{A^*(A\hat{f}^{AL}-y)+\mu
Dv|v\in\mR^d:v_i=\mbox{sgn}(D_i^*\hat{f}^{AL}) \\ \mbox{ if }
D_i^*\hat{f}^{AL}\neq 0 \mbox{ and } |v_i|\leq 1 \mbox{otherwise}
\right\}.$$ Hence there exists $v\in \mR^d$ such that
$\|v\|_{\infty}\leq 1$ satisfying
$$A^*(A\hat{f}^{AL}-y)+\mu Dv=0.$$ Now we get
$$\|D^*A^*(A\hat{f}^{AL}-y)\|_{\infty}=\mu \|D^*Dv\|_{\infty}\leq \mu \|D^*D\|_{\infty,\infty}=\mu \|D^*D\|_{1,1}.$$

Since $\hat{f}^{AL}$ is the minimizer to (\ref{ALASSO}), we have
$$\frac{1}{2}\|A\hat{f}^{AL}-y\|_2^2+\mu\|D^*\hat{f}^{AL}\|_1\leq
\frac{1}{2}\|(A{f}-y)\|_2^2+\mu\|D^*{f}\|_1.$$ Plug in $y=Af+z$ and
rearrange terms to give
\bea&&\frac{1}{2}\|Ah\|_2^2+\mu\|D^*\hat{f}^{AL}\|_1\leq \la
Ah,z\ra+\mu\|D^*{f}\|_1 .\eea From the definition of tight frame,
and then by using the holder inequality and the assumption
$\|D^*A^*z\|_{\infty}\leq \mu/2$, we have \bea\la
Ah,z\ra+\mu\|D^*{f}\|_1 =\la D^*h,D^*A^*z\ra+\mu\|D^*{f}\|_1&\leq&
\|D^*h\|_1\|D^*A^*z\|_{\infty}+\mu\|D^*{f}\|_1\\&\leq& \mu/2
\|D^*h\|_1+\mu\|D^*{f}\|_1.\eea It thus follows that
\bea\mu\|D^*\hat{f}^{AL}\|_1&\leq&\frac{1}{2}\|Ah\|_2^2+\mu\|D^*\hat{f}^{AL}\|_1\leq
 \mu/2\|D^*h\|_1+\mu\|D^*{f}\|_1.\eea This gives
 \bea\|D^*\hat{f}^{AL}\|_1\leq
 \|D^*h\|_1/2+\|D^*{f}\|_1.\eea Now a similar argument as that for
 (\ref{CC}) leads to
\be\label{e12}\|D^*_{T^c}h\|_1\leq3\|D^*_Th\|_1+4\|D^*_{T^c}f\|_1.\ee

Now we sketch the important steps of the proof. Similar to
(\ref{DAAH}), we have
$$\|D^*A^*Ah\|_{\infty}\leq\|D^*A^*(Af-y)\|_{\infty}+\|D^*A^*(A\hat{f}^{AL}-y)\|_{\infty}\leq c_0\mu,$$ where $c_0=1/2+\|D^*D\|_{1,1}.$
With the above inequality, a similar argument as that for (\ref{e3})
gives \be\label{e13}\|D^*_{T_{01}}h\|_2\leq
\frac{c_0\mu\sqrt{2s}+s^{-1/2}\delta_{3s}\|D_{T^c}^*h\|_1}{1-\delta_{3s}}.\ee
It thus follows that $$\|D^*_{T}h\|_1\leq
\sqrt{s}\|D^*_{T}h\|_2\leq\sqrt{s}\|D^*_{T_{01}}h\|_2\leq
\frac{\sqrt{2}c_0\mu
s+\delta_{3s}\|D_{T^c}^*h\|_1}{1-\delta_{3s}}.$$ Substituting the
above inequality to (\ref{e12}) and by an easy calculation, we can
obtain
\be\label{e14}\|D^*_{T^c}h\|_1\leq\frac{4(1-\delta_{3s})\|D^*_{T^c}f\|_1+3\sqrt{2}c_0\mu
s}{1-4\delta_{3s}}.\ee Using (\ref{e13}), (\ref{e2}) and then
applying (\ref{e13}), we get \bea\|h\|_2=\|D^*h\|_2\leq
\|D_{T_{01}}^*h\|_2+\sum_{j\geq2}\|D_{T_j}^*h\|_2\leq
\frac{c_0\mu\sqrt{2s}+s^{-1/2}\|D_{T^c}^*h\|_1}{1-\delta_{3s}} \leq
\frac{4\sqrt{2s}c_0\mu
}{1-4\delta_{3s}}+\frac{4\|D^*_{T^c}f\|_1}{(1-4\delta_{4s})\sqrt{s}},\eea
which leads to the result.

Repeating the above argument for each $1\leq k<s$, one can finish
the proof.
\end{proof}

\subsection{Proof of Theorem \ref{thm7}} We introduce the
following result, see \cite[Lemma 1]{LDB09}. As shown in
\cite{LDB09}, such results can be extended with different constants
to the more general class of Sub-Gaussian matrices.
\begin{lemma}\label{lemma6}
  Let $A$ be an $m\times n$ matrix with elements $a_{ij}$ drawn
  i.i.d according to $N(0,1/m)$ and let $\Phi=[A,I]$. Then for every
  $v\in\mR^{m+n},$ \be\label{e11}\mathbb{P}\left(\big|\|\Phi v\|_2^2-\|v\|_2^2\big|\geq2\delta\|v\|_2^2\right)\leq
3{e}^{- m\delta^2/8},\quad \delta\in(0,1).\ee
\end{lemma}

\begin{proof}[Proof of Theorem \ref{thm7}]
  Under the assumptions of the theorem, by Lemma \ref{lemma6}, we have that for every
  $v\in\mR^{m+n},$ (\ref{e11}) holds. Using a standard covering argument as in \cite{BDDW08}
(also \cite{RSV08}), one can prove that with probability exceeding
$1-3e^{-C_2m}$, $\Phi$ satisfies the $W$-RIP of order $s+s'$ with
constant $\delta$. Then, the conclusions follow from Theorem
\ref{thm1}, Theorem \ref{thm3}, (\ref{result(1)}) and that
\bea\|W^*u-(W^*u)_{[s+s']}\|_1\leq\|D^*f-(D^*f)_{[s]}\|_1+\|\Omega^*e-(\Omega^*e)_{[s']}\|_1.\eea
\end{proof}

\bibliographystyle{unsrt}

\end{document}